\documentclass[10pt]{iopart}   
\usepackage{iopams}  
\usepackage{epsf}  
\usepackage{times}  
 \usepackage{graphicx}
 \usepackage{color}

\eqnobysec

\usepackage{fancybox}

\begin{document}

\title{Intertwining symmetry algebras of quantum superintegrable systems 
on the hyperboloid}

\author{J A Calzada$^1$, \c{S} Kuru$^2$,
J Negro$^3$ and M A del Olmo$^3$}

\address{$^1$Departamento de Matem\'atica Aplicada, Escuela Superior de  Ingenieros Industriales, 
Universidad de Valladolid, 47011 Valladolid, Spain}

\address{$^2$Department of Physics, Faculty of Science, Ankara University,
06100 Ankara, Turkey}
\address{$^3$Departamento de F\'isica Te\'orica, At\'omica y Optica,
Facultad de Ciencias,
Universidad de Valladolid, 47011 Valladolid, Spain}

\ead{juacal@eis.uva.es, kuru@science.ankara.edu.tr, jnegro@fta.uva.es, olmo@fta.uva.es}

\begin{abstract}
A class of quantum superintegrable Hamiltonians defined on a two-dimensional
hyperboloid is considered together with a set of intertwining
operators connecting them. It is shown that such intertwining
operators close a $su(2,1)$ Lie algebra and determine the
Hamiltonians through the Casimir operators. 
By means of discrete symmetries a broader set of operators is
obtained closing a $so(4,2)$ algebra.
The physical states corresponding to the discrete spectrum 
of bound states as well as the degeneration are 
characterized in terms of unitary representations of 
$su(2,1)$ and $so(4,2)$.

\end{abstract}

\section{Introduction}

In this work we will consider a quantum superintegrable system 
living in a
two-dimensional hyperboloid of two-sheets. Although this system is
well known in the literature \cite{winternitz}-\cite{contracciones}
and can be dealt with standard procedures
\cite{ranada}-\cite{evans}, it will be studied here under a
different point of view based on the properties of intertwining
operators (IO), a form of Darboux transformations \cite{darboux}. We will see how
this approach can give a simple explanation of the main features of
this physical system. The intertwining operators and integrable Hamiltonians
have been studied in previous references
\cite{kuru1}-\cite{fernandez}, but we will supply here a thorough
non-trivial application by means of this example. Besides, there are
several points of interest for the specific case here considered because of the non-compact character.

The intertwining operators are first order differential operators
connecting different Hamiltonians in the same class (called hierarchy)
and they are associated to separable coordinates of the Hamiltonians.
We will obtain just a complete set of such intertwining operators,
in the sense that any of the Hamiltonians of the hierarchy can be
expressed in terms of these operators.

In our case the initial IO's close an algebraic structure which is the
non-compact Lie algebra $su(2,1)$
(see \cite{olmo106} for a compact case). In a second step we will
get a larger $so(4,2)$ Lie algebra of operators.
This structure allows us to characterize
the discrete spectrum and the corresponding eigenfunctions of the system by means of  (infinite dimensional) irreducible unitary representations (iur).
The construction of such representations, as it is known, is not so
standard as for compact Lie algebras.
We will compute the ground state
and characterize the  representation space
of the wave-functions which share the same energy. Notice that these systems include also a continuum spectrum, but we will not go into this point here.

The organization of the paper is as follows. Section 2 introduces 
the superintegrable Hamiltonians and Section 3 shows how to
build the IO's connecting hierarchies of these kind of Hamiltonians. 
In Section 4 it is seen that these operators
close a $su(2,1)$ algebra. The Hamiltonians are related to the
Casimirs of such an algebra, while the discrete spectrum of the
Hamiltonians is related to unitary representations (iur's) of
$su(2,1)$. Next, in Section 5 a broader class of IO's is 
defined leading to the $so(4,2)$ Lie algebra, and it is shown
how this new structure helps to understand better the
Hamiltonians in the new hierarchies. Finally, some remarks and
conclusions in Section 6 will end
the paper.

\section{Parametrizations of the two-sheet hyperboloid}\label{parametrizations}

Let us consider the two-dimensional two-sheet hyperboloid
$s_{0}^2+s_{1}^2-s_{2}^2=-1$, where
we define the following Hamiltonian
\begin{equation}\label{hamiltonian}
H_{\ell}=J_{2}^2-J_{1}^2-J_{0}^2-
\frac{l_{2}^2-\frac{1}{4}}{s_{2}^2}+\frac{l_{1}^2-\frac{1}{4}}{s_{1}^2}+\frac{l_{0}^2-\frac{1}{4}}{s_{0}^2},
\end{equation}
where $\ell=(l_{0},\,l_{1},\,l_{2}) \in \mathbb{R}^{3}$, and
the differential operators
\begin{equation}\label{diffgenerators}
J_0=s_1\partial_2+ s_2\partial_1,\quad
J_1=s_2\partial_0+ s_0\partial_2,\quad
J_2=s_0\partial_1- s_1\partial_0,
\end{equation}
constitute a realization of the $so(2,1)$ Lie algebra with Lie commutators
\[
[J_{0},J_{1}]=-J_{2}, \qquad 
[J_{2},J_{0}]=J_{1},\qquad 
[J_{1},J_{2}]=J_{0}.
\]
The generator $J_2$ corresponds to a rotation around the axis $s_2$,
while the generators $J_0$ and $J_1$ give pseudo-rotations (i.e., non-compact rotations) around
the axes $s_0$ and $s_1$, respectively.
The Casimir operator
\[
C=J_{0}^2+J_{1}^2-J_{2}^2
\]
gives the `kinetic' part of the Hamiltonian.

We can parametrize the hyperbolic surface by means of the `analogue' of the spherical coordinates
\begin{equation}\label{sa}
s_{0}={\sinh\xi}\,\cos{\theta},\quad
s_{1}={\sinh\xi}\,\sin{\theta},\quad
s_{2}={\cosh{\xi}},
\end{equation}
where $0\leq\theta<2\pi$ and $0\leq\xi<\infty$. In these coordinates,  the infinitesimal generators (\ref{diffgenerators}) take the following expressions
\begin{equation}\label{ja}
\fl
J_{0}=\sin{\theta}\,\partial_{\xi}+\cos{\theta}\,{\coth\xi}
\,\partial_{\theta} ,\quad
J_{1}=\cos{\theta}\,\partial_{\xi}-\sin{\theta}\,{\coth\xi}
\,\partial_{\theta} ,\quad
J_{2}=\partial_{\theta}\, .
\end{equation}
It is easy to check that the generators $J_i$, $i=1,2,3$, are anti-Hermitian inside the space of square-integrable 
functions with the invariant measure
$d\mu(\theta,\xi)= \sinh\xi\, d\theta d\xi$.
Using the coordinates (\ref{sa}), the Hamiltonian 
(\ref{hamiltonian}) has the expression
\begin{equation}\label{ha}
\fl
H_{\ell}=-\partial_{\xi}^2-{\coth\xi}\,\partial_{\xi}
-\frac{l_{2}^2-\frac{1}{4}}{\cosh^{2}\xi}
+\frac{1}{\sinh^{2}\xi}\left[-\partial_{\theta}^2+\frac{l_{1}^2
-\frac{1}{4}}{\sin^{2}\theta}+\frac{l_{0}^2
-\frac{1}{4}}{\cos^{2}\theta}\right].
\end{equation}
Therefore, $H_{\ell}$  can be separated in the variables $\xi$ and
$\theta$. Choosing its eigenfunctions $\Phi_{\ell}$, $H_{\ell}\Phi_{\ell}=E\Phi_{\ell}$, in the form
\begin{equation}\label{wave}
\Phi_{\ell}(\theta,\xi)=f(\theta)\,g(\xi),
\end{equation}
we get the separated equations
\begin{equation}\label{ha1}
H_{l_{0},l_{1}}^{\theta} f(\theta)\equiv
\left[-\partial_{\theta}^2+\frac{l_{1}^2-\frac{1}{4}}{\sin^{2}\theta}
+\frac{l_{0}^2-\frac{1}{4}}{\cos^{2}\theta}\right]f(\theta)
=\alpha\,f(\theta)
\end{equation}
and
\begin{equation}\label{ha1b}
\left[-\partial_{\xi}^2-{\coth\xi}\,\partial_{\xi}
-\frac{l_{2}^2-\frac{1}{4}}{\cosh^{2}\xi}
+\frac{\alpha}{\sinh^{2}\xi}\right]g(\xi)=E\,g(\xi),
\end{equation}
where $\alpha>0$ is a separation constant.

\section{A complete set of intertwining operators}
\label{intertwiningoperators}

The second order operator
at the l.h.s. of (\ref{ha1}) in the variable $\theta$ can be factorized 
in terms of first order operators \cite{barut,quesne}
\[
H_{l_{0},l_{1}}^{\theta}=A^{+}_{l_{0},l_{1}}A^{-}_{l_{0},l_{1}}+\lambda_{l_{0},l_{1}},
\]
being
\begin{equation}\label{apm}
\fl
A_{l_{0},l_{1}}^{\pm}=\pm\partial_{\theta}-(l_{0}+1/2)\,{\tan{\theta}}
+(l_{1}+1/2)\,{\cot{\theta}},\quad
\lambda_{l_{0},l_{1}}=(1+l_{0}+l_{1})^2.
\end{equation}
The Hamiltonian can also be rewritten in terms of the triplet  
($A^{\pm}_{l_{0}-1,l_{1}-1}, \lambda_{l_{0}-1,l_{1-1}}$)
\begin{equation}\label{ah}
H_{l_{0},l_{1}}^{\theta}=
A_{l_{0}-1,l_{1}-1}^{-}A_{l_{0}-1,l_{1}-1}^{+}+\lambda_{l_{0}-1,l_{1}-1}
=A_{l_{0},l_{1}}^{+}A_{l_{0},l_{1}}^{-}
+\lambda_{l_{0},l_{1}} .
\end{equation}
In this way we get a hierarchy of Hamiltonians
\begin{equation}\label{hierarchy}
 \cdots ,H_{{l_{0}-1},{l_{1}-1}}^{\theta},H_{{l_{0}},{l_{1}}}^{\theta},H_{{l_{0}+1},{l_{1}+1}}^{\theta},
\cdots, H_{{l_{0}+n},{l_{1}+n}}^{\theta},\cdots
\end{equation}
satisfying the following recurrence relations 
\[
\begin{array}{l}
A_{l_{0}-1,l_{1}-1}^{-}H_{l_{0}-1,l_{1}-1}^{\theta}=H_{l_{0},l_{1}}^{\theta}A_{l_{0}-1,l_{1}-1}^{-},\\[1.5ex]
A_{l_{0}-1,l_{1}-1}^{+}H_{l_{0},l_{1}}^{\theta}=H_{l_{0}-1,l_{1}-1}^{\theta}A_{l_{0}-1,l_{1}-1}^{+}\, .
\end{array}
\]
Hence, the operators   $\{ A^\pm _{l_{0}+n,l_{1}+n}\}_{n\in\mathbb{Z}}$ are intertwining operators and they act as transformations between
the eigenfunctions of the Hamiltonians in the hierarchy 
(\ref{hierarchy}),
\[
A_{l_{0}-1,l_{1}-1}^{-}:f_{l_{0}-1,l_{1}-1}\rightarrow
f_{l_{0},l_{1}},\qquad
A_{l_{0}-1,l_{1}-1}^{+}:f_{l_{0},l_{1}}\rightarrow
f_{l_{0}-1,l_{1}-1},
\]
where the subindex refers to the corresponding Hamiltonian.
We can define new operators in terms of $A_{l_{0},l_{1}}^{\pm}$,
together with a diagonal operator  
$A_{l_{0},l_{1}}=(l_{0}+l_{1})\mathbb{I}$, acting in the following way
in the space of eigenfunctions
\begin{equation}\label{noindex}
\fl
\hat{A}^{-} f_{l_{0},l_{1}}:=\frac{1}{2}\,A_{l_{0},l_{1}}^{-}
\,f_{l_{0},l_{1}},\ 
\hat{A}^{+} f_{l_{0},l_{1}}:=\frac{1}{2}\,A_{l_{0},l_{1}}^{+}
\,f_{l_{0},l_{1}},\ 
\hat{A}\,f_{l_{0},l_{1}}:=-\frac{1}{2}\,(l_{0}+l_{1})\,f_{l_{0},l_{1}}.
\end{equation}
It can be shown from (\ref{ah}) that $\{\hat{A}^{-}, \hat{A}^{+},
\hat{A}\}$ satisfy the commutation
relations of a $su(2)$ Lie algebra, i.e.,
\begin{equation}\label{a}
[\hat{A}^{-}, \hat{A}^{+}]=-2\,\hat{A},\qquad [\hat{A},
\hat{A}^{\pm}]=\pm \hat{A}^{\pm}.
\end{equation}
The `fundamental' states, $f_{l_{0},l_{1}}^{0}$, of the $su(2)$ representations
are determined by the relation
$\hat A^{-}\,f_{l_{0},l_{1}}^{0}(\theta)=0$. They are
\[
f_{l_{0},l_{1}}^{0}(\theta)=N\,(\cos{\theta})^{l_{0}+1/2}\,(\sin{\theta})^{l_{1}+1/2}, 
\]
where $N$ is a normalization constant. These functions are regular
and square-integrable when 
\begin{equation}\label{des1}
l_0,l_1\geq -1/2\, .
\end{equation}
Since
$\hat{A}\,f_{l_{0},l_{1}}\equiv
-\frac{1}{2}\,(l_{0}+l_{1})\,f_{l_{0},l_{1}}$,
then the label of the $j$-representation is 
$j=\frac{1}{2}\,(l_{0}+l_{1})$ and the dimension of the iur will be $2j+1=l_{0}+l_{1}+1$.

Now, observe that because the IO's $A^{\pm}_{l_{0},l_{1}}$ depend
only on the $\theta$-variable, they can act also as IO's of the
total Hamiltonians $H_{\ell}$ (\ref{ha}) and its global
eigenfunctions $\Phi_{\ell}$ (\ref{wave}), leaving the parameter
$l_2$ unchanged (in this framework we will use three-fold indexes)
\[
A_{\ell'}^{-}H_{\ell'}=H_{\ell}A_{\ell'}^{-},\qquad
A_{\ell'}^{+}H_{\ell}=H_{\ell'}A_{\ell'}^{+},
\]
where $\ell=(l_{0},l_{1},l_2)$ and $\ell'=(l_{0}-1,l_{1}-1,l_2)$.
In this sense, many of the above relations can be straightforwardly 
extended under this global point of view.

\subsection{Second set of pseudo-spherical coordinates}

A second coordinate set is obtained from the non-compact rotations about the axes
$s_{0}$ and $s_{1}$, respectively. In this way we obtain the
following parametrization of the hyperboloid
\begin{equation}\label{sb}
s_{0}={\cosh\psi}\,{\sinh\chi},\qquad
s_{1}={\sinh\psi},\qquad
s_{2}={\cosh\psi}\,{\cosh\chi}\, .
\end{equation}
The expressions of the $so(2,1)$ generators in these coordinates
are
\[
\fl
J_{0}=-{\tanh{\psi}}\,{\sinh{\chi}}\partial_{\chi}
+{\cosh{\chi}}\;\partial_{\psi},\quad
J_{1}=\partial_{\chi} ,\quad
J_{2}={\sinh{\chi}}\partial_{\psi}-{\tanh{\psi}}
\,{\cosh{\chi}}\;\partial_{\chi}\, 
\]
and the explicit expression of the Hamiltonian (\ref{hamiltonian}) is now
\[
\fl
H_{\ell}=-\partial_{\psi}^2-{\tanh\psi}\,\partial_{\psi}
+\frac{l_{1}^2-\frac{1}{4}}{\sinh^{2}\psi}
+\frac{1}{\cosh^{2}\psi}\left[-\partial_{\chi}^2+\frac{l_{0}^2
-\frac{1}{4}}{\sinh^{2}\chi}-\frac{l_{2}^2-\frac{1}{4}}{\cosh^{2}\chi}\right].
\]
This Hamiltonian can be separated in the variables $\psi$ and $\chi$
considering the eigenfunctions $\Phi$ of $H_\ell$  ($H_{\ell}\,\Phi=E\,\Phi$) as $\Phi(\chi,\psi)=f(\chi)\,g(\psi)$. We obtain the two folllowing (separated) equations
\begin{equation}\label{hb1}
H_{l_{0},l_{2}}^{\chi}f(\chi)\equiv
\left[-\partial_{\chi}^2+\frac{l_{0}^2
-\frac{1}{4}}{\sinh^{2}\chi}-\frac{l_{2}^2
-\frac{1}{4}}{\cosh^{2}\chi}\right]f(\chi)
=\alpha\,f(\chi),
\end{equation}
\[
\left[-\partial_{\psi}^2
-{\tanh\psi}\,\partial_{\psi}+\frac{l_{1}^2
-\frac{1}{4}}{\sinh^{2}\psi}+\frac{\alpha}{\cosh^{2}\psi}\right]g(\psi)
=E\,g(\psi),
\]
with $\alpha$ a separation constant. The second order operator 
in the variable $\chi$ at
the l.h.s.\ of (\ref{hb1}) can be factorized as a product of
first order operators
\begin{equation}\label{hbb1}
H_{l_{0},l_{2}}^{\chi}=B^{+}_{l_{0},l_{2}}B^{-}_{l_{0},l_{2}}
+\lambda_{l_{0},l_{2}}=
B^{-}_{l_{0}-1,l_{2}-1}B^{+}_{l_{0}-1,l_{2}-1}+\lambda_{l_{0}-1,l_{2}-1},
\end{equation}
being
\begin{equation}
\fl
\label{bpm}
B_{l_{0},l_{2}}^{\pm}=\pm\partial_{\chi}+(l_{2}+1/2)\,
\tanh{\chi}+(l_{0}+1/2)
\,{\coth{\chi}}, \quad\lambda_{l_{0},l_{2}}=-(1+l_{0}+l_{2})^2.
\end{equation}
In this case the intertwining relations take the form
\[
\begin{array}{l}
B^{-}_{l_{0}-1,l_{2}-1}H_{l_{0}-1,l_{2}-1}^{\chi}=H_{l_{0},l_{2}}^{\chi}
B^{-}_{l_{0}-1,l_{2}-1},\\[1.5ex]
B^{+}_{l_{0}-1,l_{2}-1}H_{l_{0},l_{2}}^{\chi}=
H_{l_{0}-1,l_{2}-1}^{\chi}B^{+}_{l_{0}-1,l_{2}-1},
\end{array}
\]
and imply that these operators $B^{\pm}$ connect eigenfunctions in
the following way
\[
B^{-}_{l_{0}-1,l_{2}-1}:f_{l_{0}-1,l_{2}-1}\rightarrow
f_{l_{0},l_{2}},\qquad
B^{+}_{l_{0}-1,l_{2}-1}:f_{l_{0},l_{2}}\rightarrow
f_{l_{0}-1,l_{2}-1}\,.
\]
The operators $B^{\pm}_{l_{0},l_{2}}$ can be expressed in terms of
$\xi$ and $\theta$ using relations (\ref{sa}) and (\ref{sb})
\[
B^{\pm}_{l_{0},l_{2}}=\pm J_{1}+(l_{2}+1/2)
\,{\tanh{\xi}}\,\cos{\theta}+(l_{0}+1/2)\,{\coth{\xi}}\,\sec{\theta},
\]
where $J_{1}$ is given by (\ref{ja}). We define new free-index
operators in the following way
\[
\fl
\hat{B}^{-}\,f_{l_{0},l_{2}}:=\frac{1}{2}\,B_{l_{0},l_{2}}^{-}
\,f_{l_{0},l_{2}},\quad
\hat{B}^{+}\,f_{l_{0},l_{2}}:=\frac{1}{2}\,B_{l_{0},l_{2}}^{+}
\,f_{l_{0},l_{2}},\quad
\hat{B}\,f_{l_{0},l_{2}}:=-\frac{1}{2}\,(l_{0}+l_{2})\,f_{l_{0},l_{2}},
\]
and, having in mind the expressions (\ref{hbb1}) and (\ref{bpm}), we can prove than they close the
$su(1,1)$ Lie algebra
\begin{equation}\label{b}
[\hat{B}^{-}, \hat{B}^{+}]=2\,\hat{B},\qquad 
[\hat{B},\hat{B}^{\pm}]=\pm \hat{B}^{\pm}.
\end{equation}
Since the Lie algebra $su(1,1)$ is non-compact, its 
iur's are infinite dimensional. In particular, we will
be interested in the discrete series, that is, 
in those having a fundamental state annihilated by
the lowering operator, i.e., 
$B^{-} \,f^0_{l_{0},l_{2}}=0$. The explicit 
expression of these states is
\begin{equation}
f^0_{l_{0},l_{2}}(\chi)
= N (\cosh \chi)^{l_2+1/2}(\sinh \chi)^{l_0+1/2}
\end{equation}
where $N$ is a normalization constant. 
In order to have a regular 
and square-integrable function we must have
\begin{equation}\label{des2}
l_0\geq -1/2,\quad -k_1\equiv l_0+l_2<-1 \, .
\end{equation}
Since
$
\hat{B}\,f^0_{l_{0},l_{2}}=
-\frac{1}{2}\,(l_{0}+l_{2})\,f^0_{l_{0},l_{2}}
$, we can say that the lowest weight of this unitary $su(1,1)$ infinite representation is $j_1'= k_1/2>1/2$.

The IO's $\hat B^{\pm}$ can be considered also as intertwining operators 
of the Hamiltonians $H_{\ell}$ linking their eigenfunctions 
$\Phi_{\ell}$, similarly to the IO's $\hat A^{\pm}$ described before (in
this situation we will also use three-fold indexes but now with $l_1$ 
remaining unchanged).

\subsection{Third set of pseudo-spherical coordinates}

A third set of coordinates is obtained from the noncompact rotations about the axes
$s_{1}$ and $s_{0}$, respectively. They  give rise to the parametrization
\begin{equation}\label{sc}
s_{0}={\sinh\phi},\quad 
s_{1}={\cosh\phi}\,{\sinh\beta},\quad 
s_{2}={\cosh\phi}\,{\cosh\beta},
\end{equation}
and the generators have the expressions
\[
\fl
J_{0}=\partial_{\beta},\quad
J_{1}={\cosh{\beta}}\;\partial_{\phi}-{\tanh{\phi}}\,{\sinh{\beta}}
\;\partial_{\beta} ,\quad
J_{2}=-{\sinh{\beta}}\;\partial_{\phi}
+{\tanh{\phi}}\,{\cosh{\beta}}\;\partial_{\beta} .
\]
Now, the Hamiltonian takes the form
\[
\fl
H_{\ell}=-\partial_{\phi}^2-{\tanh\phi}\,\partial_{\phi}
+\frac{l_{0}^2-\frac{1}{4}}{\sinh^{2}\phi}
+\frac{1}{\cosh^{2}\phi}\left[-\partial_{\beta}^2+\frac{l_{1}^2
-\frac{1}{4}}{\sinh^{2}\beta}-\frac{l_{2}^2
-\frac{1}{4}}{\cosh^{2}\beta}\right],
\]
and can be separated in the variables $\phi,\beta$ in terms of  its eigenfunctions 
$\Phi$ $(H_{\ell}\,\Phi=E\,\Phi)$ such that 
$\Phi(\beta,\phi)=f(\beta)\,g(\phi)$ in the following way
\begin{equation}\label{hc1}
H_{l_{1},l_{2}}^{\beta}f(\beta)\equiv
\left[-\partial_{\beta}^2+\frac{l_{1}^2
-\frac{1}{4}}{\sinh^{2}\beta}-\frac{l_{2}^2
-\frac{1}{4}}{\cosh^{2}\beta}\right]f(\beta)=\alpha\,f(\beta),
\end{equation}
\[
\left[-\partial_{\phi}^2
-{\tanh\phi}\,\partial_{\phi}+\frac{l_{0}^2-\frac{1}{4}}{\sinh^{2}\phi}
+\frac{\alpha}{\cosh^{2}\phi}\right]g(\phi)=
E\,g(\phi), 
\]
with the separation constant $\alpha$. The second order operator in 
$\beta$
at the l.h.s.\ of expression (\ref{hc1}) can be factorized as a product
of first order operators
\[
H_{l_{1},l_{2}}^{\beta}={C}^{+}_{l_{1},l_{2}}{C}^{-}_{l_{1},l_{2}}+{\lambda}_{l_{1},l_{2}}
={C}^{-}_{l_{1}+1,l_{2}-1}{C}^{+}_{l_{1}+1,l_{2}-1}+
{\lambda}_{l_{1}+1,l_{2}-1},
\]
being
\[
\fl
C_{l_{1},l_{2}}^{\pm}=
\pm\partial_{\beta}+(l_{2}+1/2)\,{\tanh{\beta}}+(-l_{1}+1/2)
\,{\coth{\beta}},\quad
\lambda_{l_{1},l_{2}}=-(1-l_{1}+l_{2})^2 .
\]
These operators ${C}^{\pm}_{l_{1},l_{2}}$ give rise to the intertwining relations
\[
\begin{array}{l}
{C}^{+}_{l_{1}+1,l_{2}-1}H_{l_{1},l_{2}}^{\beta}=
H_{l_{1}+1,l_{2}-1}^{\beta}C^{+}_{l_{1}+1,l_{2}-1} \\[1.5ex]
{C}^{-}_{l_{1}+1,l_{2}-1}H_{l_{1}+1,l_{2}-1}^{\beta}=
H_{l_{1},l_{2}}^{\beta}{C}^{-}_{l_{1}+1,l_{2}-1},
\end{array}
\]
which imply the connection among eigenfunctions
\[
{C}^{-}_{l_{1}+1,l_{2}-1}:f_{l_{1}+1,l_{2}-1}\rightarrow
f_{l_{1},l_{2}},\qquad
{C}^{+}_{l_{1}+1,l_{2}-1}:f_{l_{1},l_{2}}\rightarrow
f_{l_{1}+1,l_{2}-1}\,.
\]
In this case  $C^{\pm}_{l_{1},l_{2}}$ can also be expressed in
terms of $\xi$ and $\theta$ using relations (\ref{sa}) and (\ref{sc})
\[
C^{\pm}_{l_        {1},l_{2}}=\pm J_{0}+(l_{2}+1/2)
\,{\tanh{\xi}}\,\sin{\theta}+(-l_{1}+1/2)\,{\coth{\xi}}\, \csc{\theta},
\]
where $J_{0}$ is given by (\ref{ja}). Now, the new operators
are defined as
\[
\fl
\hat{C}^{-}\,f_{l_{1},l_{2}}:=\frac{1}{2}\,{C}_{l_{1},l_{2}}^{-}
\,f_{l_{1},l_{2}}\quad
\hat{C}^{+}\,f_{l_{1},l_{2}}:=\frac{1}{2}\,{C}_{l_{1},l_{2}}^{+}
\,f_{l_{1},l_{2}}\quad
\hat{C}\,f_{l_{1},l_{2}}:=-\frac{1}{2}\,(l_{2}-l_{1})f_{l_{1},l_{2}}.
\]
and satisfy the commutation relations of the $su(1,1)$ algebra
\begin{equation}\label{c}
[\hat{C}^{-}, \hat{C}^{+}]=2\,\hat{C},\qquad 
[\hat{C}, \hat{C}^{\pm}]=\pm \hat{C}^{\pm}.
\end{equation}
The fundamental state for the $su(1,1)$ representation, in this
case, given by 
$\hat {C}^{-}\,f^0_{l_{1},l_{2}}=0$, has the expression
\begin{equation}
f^0_{l_{1},l_{2}}(\beta)=
N (\cosh \beta)^{l_2 +1/2} (\sinh \beta)^{-l_1 +1/2}
\end{equation}
where $N$ is a normalization constant. In order to get a iur from
this function, we impose it to be regular and normalizable, therefore
\begin{equation}\label{des3}
l_1\leq 1/2,\quad 
-k_2\equiv l_2-l_1< -1 \, .
\end{equation}
Since 
$
\hat{C} f^0_{l_{1},l_{2}}=
-\frac{1}{2}\,(l_{2}-l_{1})f^0_{l_{1},l_{2}}
$
the lowest weight of the iur\ is given by
$j'_2= k_2/2>1/2$.

As in the previous cases, we can consider the IO's $C^{\pm}$
as connecting global Hamiltonians $H_{\ell}$ and their
eigenfunctions, having in mind that now the parameter $l_0$ is
unaltered.

\section{Algebraic structure of the intertwining operators}

If we consider together all the IO's 
$\{\hat{A}^{\pm},\hat{A},\hat{B}^{\pm},
\hat{B},\hat{C}^{\pm},\hat{C}\}$
that have appeared in 
section~\ref{intertwiningoperators},
then, we find that they close the Lie algebra  
$su(2,1)$ since they
satisfy, besides (\ref{a}),
(\ref{b}), (\ref{c}), the following commutation relations
\begin{equation}\label{su21}
\fl
\label{abc}
\begin{array}{llll}
[\hat{A}^{+}, \hat{B}^{+}]=0\quad
&[\hat{A}^{-},\hat{B}^{-}]=0\quad
&[\hat{A}^{+},\hat{B}^{-}]=-\hat{C}^{-}\quad
&[\hat{A}^{-},\hat{B}^{+}]=\hat{C}^{+}\nonumber\\[2.ex]
[\hat{C}^{+}, \hat{B}^{+}]=0\quad
&[\hat{C}^{-}, \hat{B}^{-}]=0\quad
&[\hat{C}^{+}, \hat{A}^{+}]=-\hat{B}^{+}\quad
&[\hat{C}^{-},\hat{A}^{-}]=\hat{B}^{-}\nonumber\\[2.ex]
[\hat{C}^{+}, \hat{B}^{-}]=-\hat{A}^{-}\quad
&[\hat{C}^{-}, \hat{B}^{+}]=\hat{A}^{+}\quad
&[\hat{C}^{+},\hat{A}^{-}]=0\quad
&[\hat{C}^{-},\hat{A}^{+}]=0\nonumber\\[2.ex]
[\hat{A}, \hat{B}^{+}]=\frac{1}{2} \hat{B}^{+}\quad
&[\hat{A},\hat{B}^{-}]=-\frac{1}{2} \hat{B}^{-}\quad
&[\hat{B},\hat{A}^{+}]=\frac{1}{2} \hat{A}^{+}\quad
&[\hat{B},\hat{A}^{-}]=-\frac{1}{2} \hat{A}^{-}\nonumber\\[2.ex]
[\hat{C}, \hat{B}^{+}]=\frac{1}{2} \hat{B}^{+}\quad
&[\hat{C},\hat{B}^{-}]=-\frac{1}{2} \hat{B}^{-}\quad
&[\hat{C},\hat{A}^{+}]=-\frac{1}{2} \hat{A}^{+}\quad
&[\hat{C},\hat{A}^{-}]=\frac{1}{2} \hat{A}^{-}\nonumber\\[2.ex]
[\hat{A}, \hat{C}^{-}]=\frac{1}{2} \hat{C}^{-}\quad
&[\hat{A}, \hat{C}^{+}]=-\frac{1}{2} \hat{C}^{+}\quad
&[\hat{B}, \hat{C}^{-}]=-\frac{1}{2} \hat{C}^{-}\quad
&[\hat{B},\hat{C}^{+}]=\frac{1}{2} \hat{C}^{+}\nonumber\\[2.ex]
[\hat{A}, \hat{B}]=0\quad
&[\hat{A}, \hat{C}]=0\quad
&[\hat{B}, \hat{C}]=0\, .& \qquad\nonumber
\end{array}
\end{equation}
Obviously  $su(2,1)$  includes as
subalgebras the Lie algebras $su(2)$ and $su(1,1)$ defined in the previous section \ref{intertwiningoperators}. The
second order Casimir operator of $su(2,1)$  can be written as follows
\begin{equation}
\label{cas}
\fl
{\cal C} = \hat A^+\hat A^- - \hat B^+\hat B^- - \hat C^+\hat C^- +
\frac23\left(
\hat A^2+  \hat B^2 +   \hat C^2\right) -
(\hat A+\hat B+\hat C)\, .
\end{equation}
It is worthy noticing  that in our differential realization we have
$\hat A - \hat B + \hat C =0$,
and that there is another generator
\begin{equation}\label{cas2}
{\cal C}'= l_1+l_2-l_0
\end{equation}
commuting with the rest of generators. Hence, adding this new generator ${\cal C}'$ to the other ones we get the Lie algebra $u(2,1)$.   

The eigenfunctions of the Hamiltonians $H_{\ell}$
that have the same energy support unitary representations
of $su(2,1)$ characterized by a value of
${\cal C}$ and other of  ${\cal C}'$. In fact, we can show that
\begin{equation}\label{hcc}
H_{\ell} = -4\, {\cal C}+ \frac13\,  {{\cal C}'}^2 - \frac{15}{4}.
\end{equation}
These representations can be obtained, as usual, starting with
a fundamental state simultaneously annihilated by the lowering operators 
$\hat A^-, \hat C^-$ and  $\hat B^-$
\begin{equation}\label{equation1}
 A^-_{\ell}\Phi^0_{\ell} = C^-_{\ell} \Phi^0_{\ell} =
B^-_{\ell} \Phi^0_{\ell} =0\, .
\end{equation}
Solving equations (\ref{equation1}) we find 
\begin{equation}\label{fs}
\Phi^0_{\ell}(\xi,\theta) =
N (\cos \theta)^{l_0+1/2}
(\sin \theta)^{1/2} (\cosh \xi)^{l_2+1/2}(\sinh \xi)^{l_0+1},
\end{equation}
where $\ell= (l_0, 0, l_2)$. 
From the inequalities (\ref{des1}) and (\ref{des3}) the parameters
of $\Phi^0_{\ell}$ must satisfy $(l_0+l_2)<-1$ and $l_0\geq -1/2$.
In this particular case to guarantee the normalization of 
$\Phi^0_{\ell}$ using the invariant measure 
we must impose $(l_0+l_2)<-5/2$.
Thus, the above state supports also 
iur's of the subalgebras $su(2)$ (generated by $\hat A^{\pm}$) with
the weight $j=l_0/2$ and $su(1,1)$ (generated by $\hat C^{\pm}$) 
with $j'_2=-l_2/2$.

The energies of the fundamental states of the form (\ref{fs}) are 
obtained by applying $H_{\ell}$ as given in (\ref{hcc}) 
taking into account 
with the expressions for the Casimir operators (\ref{cas}) and 
(\ref{cas2}),
\begin{equation}\label{energy}
H_{\ell} \Phi^0_{\ell} = - (l_0+l_2+3/2)(l_0+l_2+5/2) \Phi^0_{\ell}
\equiv E^0_{\ell} \Phi^0_{\ell} \, .
\end{equation}
From $\Phi^0_{\ell}$ we can get the rest of eigenfunctions in the
$su(2,1)$ representation using the raising operators 
$\hat A^+,\hat B^+$ 
and $\hat C^+$, all of them sharing the same energy eigenvalue 
$E^0_{\ell}$ (\ref{energy}).

\begin{figure}[h]
\centering
\includegraphics[scale=0.5]{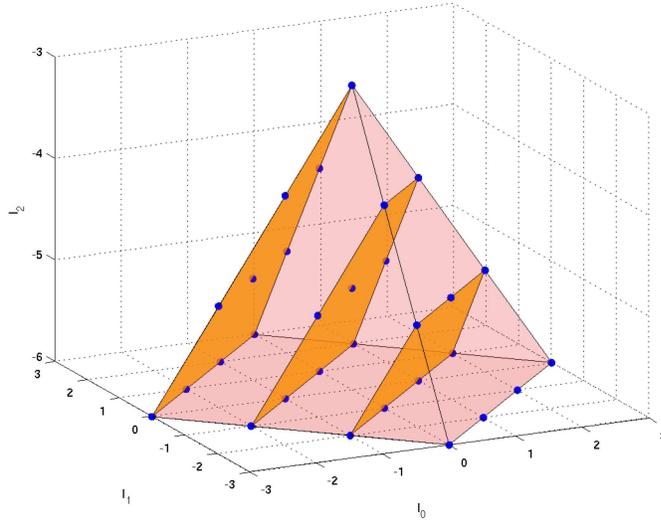}
\caption{\small The states of iur's of $su(2,1)$
represented by points in the three dark planes corresponding
to $\Phi^0_{\ell}$ with $\ell=(0,0,-3)$,
$\ell=(1,0,-4)$ and $\ell=(2,0,-5)$. All of them share
the same energy.
} \label{fig1}
\end{figure}

Since the expression (\ref{energy}) for $E^0_{\ell}$  depends on
$l_0+l_2$, it means that states in the family of iur's derived from fundamental states (\ref{fs}) such
that have the same value of $l_0+l_2$ will also have the same
energy eigenvalue. It is worth to remark that the energy in 
(\ref{energy}) corresponding to bound states is negative, 
which is consistent with
the expressions (\ref{ha}) and (\ref{ha1b}) for the Hamiltonians,
and that the set of such bound states for each Hamiltonian is finite.

In Fig.~\ref{fig1}, by means of an example, 
we represent the states of some iur's by points
$(l_0,l_1,l_2)\in \mathbb{R}^3$ linked to the ground state
$\Phi^0_{\ell}$, represented by the point $(l_0,0,l_2)$, through
the raising operators  $\hat A^+,\hat C^+$. 
The points belonging to a iur\
are in a 2-dimensional plane (corresponding to the particular value of ${\cal C}'$), and other iur's are described by points in 
parallel 2D  planes. These parallel planes are closed inside a 
tetrahedral unbounded pyramid whose basis extends towards $l_2\to -\infty$.

As in the case of $su(3)$ representations\cite{olmo106}, in the above 
$su(2,1)$ iur's we have some points (in the parameter space) which
are degenerated, that is, they correspond to an eigen-space whose 
dimension is greater than one. For example, let us consider first the  representation
based on the fundamental state $\Phi^0_{\ell}$ with values
$\ell=(0,0,-3)$. From this state we can build a iur made of points
in a triangle, where each point represents a non-degenerate 1-dimensional (1D) eigenspace. 
Now, consider the iur corresponding to the ground
state with $\ell_0=(1,0,-4)$, which has eigenstates with the same
energy $E= -(-3+3/2)(-3+5/2)$ as the previous one 
(they have the same value of $l_0+l_2=-3$).
Now the eigenstates corresponding to $\ell_1=(0,0,-5)$, inside this representation, can be obtained in two ways: 
\begin{equation}
\Phi_{\ell_1} = \hat C^+ \hat A^+ \Phi_{\ell_0},\quad
\tilde\Phi_{\ell_1} = \hat A^+ \hat C^+ \Phi_{\ell_0} \, .
\end{equation}
It can be shown that these states are independent and that they span 
the 2D eigenspace of
the corresponding Hamiltonian $H_{\ell_1}$ for that eigenvalue.

\begin{figure}[h]
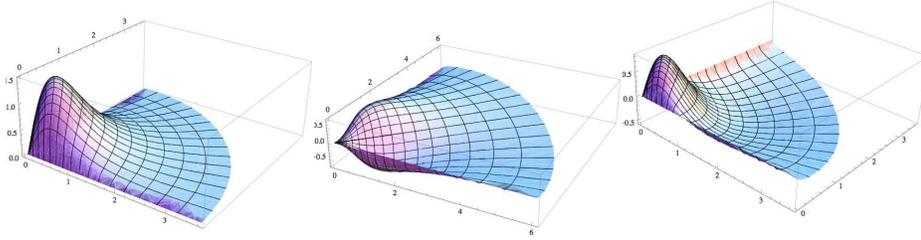

\centering
\includegraphics[scale=0.4]{fund5.epsf}
\hskip-0.05cm
\includegraphics[scale=0.4]{fig2a.epsf}
\hskip-0.05cm
\includegraphics[scale=0.4]{fig2b.epsf}
\caption{\small Plot of the orthogonal normalized eigenfunctions
$\Phi(\theta,\xi)$, $0<\theta<\pi/2$, $0<\xi<\infty$,
of $H_{\ell}$, for $\ell=(0,0,-5)$ corresponding 
to the ground state (left) and 
two independent states generating the eigenspace of
the second (and last) excited energy level.} \label{fig2}
\end{figure}

Remark that the ground state for the Hamiltonian $H_{\ell_1}$,
$\ell_1=(0,0,-5)$, is given by the wavefunction (\ref{fs}) and its
ground  energy is $E^0_{\ell_1}= - (-5+3/2)(-5+5/2)$. The plot of
the ground wavefunction and two independent excited wavefunctions
are shown in Fig.~\ref{fig2}.

Following the same pattern it can be obtained the degeneration of 
higher excited levels in the discrete spectrum: 
the $n$ excited level, when it exists,
has associated an eigenspace with dimension $n$.

\section{The complete symmetry algebra ${so(4,2)}$}

As it is explicit from its expression (\ref{hamiltonian}) the
Hamiltonian $H_{\ell}$ is invariant under reflections
\begin{equation}
\begin{array}{l}
I_0: (l_0,l_1,l_2)\to (-l_0,l_1,l_2)\\[1.5ex]
I_1: (l_0,l_1,l_2)\to (l_0,-l_1,l_2)\\[1.5ex]
I_2: (l_0,l_1,l_2)\to (l_0,l_1,-l_2) \, .
\end{array}
\end{equation}
These operators can generate, by means of conjugation, other sets of intertwining operators from the ones already defined. For instance,
\begin{equation}
I_0: \{ \hat A^{\pm},\hat A \} \to 
\{ \tilde A^{\pm}= I_0 \hat A^{\pm}I_0, \tilde A = I_0 \hat A I_0 \} 
\end{equation}
where, from (\ref{apm}) we get
\begin{equation}\label{apmt}
\fl
\tilde A_{l_{0},l_{1}}^{\pm}=\pm\partial_{\theta}-(-l_{0}+1/2)\,{\tan{\theta}}
+(l_{1}+1/2)\,{\cot{\theta}},\quad
\tilde \lambda_{l_{0},l_{1}}=(1-l_{0}+l_{1})^2
\end{equation} 
such that
\[
\tilde A_{l_{0},l_{1}}^{-}:f_{l_{0},l_{1}}\rightarrow
f_{l_{0}-1,l_{1}+1},\quad
\tilde A_{l_{0},l_{1}}^{+}:f_{l_{0-1},l_{1}+1}\rightarrow
f_{l_{0},l_{1}} \, .
\]
Thus, we can define the operators $\tilde A^\pm$ as in (\ref{noindex})
that together with $\tilde{A}\,f_{l_{0},l_{1}}:=-\frac{1}{2}\,(-l_{0}+l_{1})\,f_{l_{0},l_{1}}$ close a second $\widetilde{su}(2)$.

Other sets of operators 
$\{\tilde B^{\pm},\tilde B\}$ and 
$\{\tilde C^{\pm},\tilde C\}$, closing $\widetilde{su}(1,1)$ Lie algebras, can also be defined with the help of these reflections in
the following way (the choice is non unique)
\begin{equation}\label{tildes}
\begin{array}{l}
I_0:\{\hat A^\pm, \hat A;\  \hat B^\pm,\hat  B;\ \hat C^\pm, \hat C \} \to
\{\tilde A^\pm, \tilde A;\ \tilde B^\pm, \tilde B;\  C^\pm, C \}
\\[1.5ex]
I_1:\{\hat A^\pm, \hat A;\  \hat B^\pm,\hat  B;\ \hat C^\pm, \hat C \} \to
\{\tilde A^\mp, -\tilde A;\  B^\pm,  B;\ \tilde C^\pm, \tilde C \}
\\[1.5ex]
I_2:\{\hat A^\pm, \hat A;\  \hat B^\pm,\hat  B;\ \hat C^\pm, \hat C \} \to
\{  A^\pm,   A;\ \tilde B^\mp,-\tilde B;\ -\tilde C^\mp, -\tilde C \}
\, .
\end{array}
\end{equation}
The whole set of the operators 
\begin{equation}\label{so42}
\{\hat A^\pm,\tilde A^\pm, 
\hat B^\pm, \tilde B^\pm, \hat C^\pm,\tilde C^\pm, L_0, L_1, L_2\}
\end{equation} 
where the diagonal operators $L_i$ are defined as
\begin{equation}
L_i \Psi_\ell = l_i \Psi_\ell
\end{equation}
generate an $o(4,2)$ Lie algebra of rank three 
with commutations rules that 
can be easily derived from those of $su(2,1)$ given in 
(\ref{su21}) and the action of the reflections (\ref{tildes}).
These generators link eigenstates of the Hamiltonians $H_\ell$
with the same eigenvalue.

Now, consider a fundamental state $\Psi^0_\ell$ for the
$so(4,2)$ algebra annihilated by the lowering operators,
\begin{equation}\label{equation2}
A^-_{\ell}\Psi^0_{\ell} =\tilde A^-_{\ell}\Psi^0_{\ell}= 
C^-_{\ell} \Psi^0_{\ell} =\tilde C^-_{\ell} \Psi^0_{\ell}=
B^-_{\ell} \Psi^0_{\ell} = \tilde B^-_{\ell} \Psi^0_{\ell} = 0\, .
\end{equation}
This state should be a particular case of (\ref{fs}) invariant
also under the $l_0$-reflection:
\begin{equation}\label{fs2}
\Phi^0_{\ell}(\xi,\theta) =
N (\cos \theta)^{1/2}
(\sin \theta)^{1/2} (\cosh \xi)^{l_2+1/2}\sinh \xi,
\end{equation}
thus, it has the label $\ell=(l_0=0,l_1=0,l_2)$, where $l_2<-5/2$. This point in the parameter space, for the example of Fig.~\ref{fig1}, corresponds to the top vertex of the pyramid,
from which all the other points displayed in the figure can
be obtained with the help of raising operators. 
Such points correspond to a iur of the $so(4,2)$
algebra, including the series of iur's of the $su(2,1)$ algebra
mentioned in the previous section.

Fixed the iur\ corresponding to $\ell=(0,0,l_2)$ such that
$-7/2\leq l_2 <-5/2$, then, the points on the surface of the
associated pyramid in the parameter space correspond to 
non-degenerated ground levels of their respective Hamiltonians.
This `top' pyramid includes other `inner' pyramids, see Fig.~3, with vertexes 
$\ell_{n} = (0,0,l_2-2n)$. Each point  
on the surface of an inner pyramid associated to $\ell_n$ represents 
an $n$-excited level $(n+1)$-fold degenerated of the iur associated to 
$\ell$.

Finally, we must remark that the same set of Hamiltonians and
eigenstates can be described by `dual' representations of
$so(4,2)$ (or $su(2,1)$) by means of inverted pyramids with positive
values of $l_2$ fixing the inverted vertex.

\begin{figure}[h]
\centering
\includegraphics[scale=0.5]{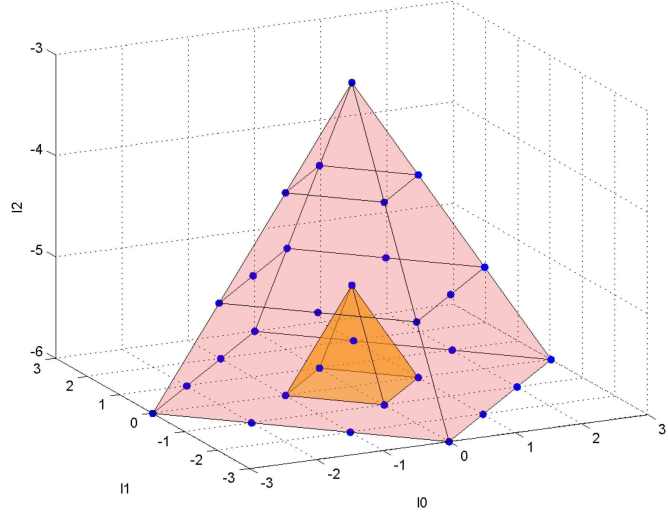}
\caption{\small It is shown two pyramids associated to the
same iur of $so(4,2)$. The exterior
with vertex $(0,0,-3)$ has points on its exterior faces
which represent non-degenerated levels. The inner one 
has exterior faces corresponding to first excited
double-degenerated levels.} \label{fig3}
\end{figure}


\section{Concluding remarks}
In this work we have built a set of intertwining operators for a superintegrable system defined on a
two-sheet hyperboloid and we have found that they close a
non-compact $su(2,1)$ Lie algebra structure. By using the 
reflections operators of the system we have implemented these
IO's obtaining an $so(4,2)$ algebra. These IO's lead to hierarchies 
of Hamiltonians described by points on planes ($su(2,1)$) or
in the 3D space ($so(4,2)$), corresponding to the rank of the
respective Lie algebra.

We have shown how these IO's can be very helpful in the characterization of the physical system by selecting
separable coordinates, determining the eigenvalues and building
eigenfunctions. We have also displayed the relation of eigenstates
and eigenvalues with unitary representations of the $su(2,1)$ and
$so(4,2)$ Lie algebras. In particular we have studied the degeneration
problem as well as the number of bound states. Remark that such a detailed study of a `non-compact' superintegrable system has not 
been realized till now, up to our knowledge.

We have restricted to iur's, but a wider analysis can be done
for hierarchies associated to representations with a not well
defined unitary character.   

The IO's can also be used to find the second order
integrals of motion for a Hamiltonian $H_{\ell}$ and their algebraic
relations, which is the usual approach to (super) integrable
systems. However, we see that it is much easier to deal directly
with the IO's, which are more elementary and simpler, than with
constants of motion.

Our program in the near future is the application of this method
to wider situations. Besides, in principle,
we can also adapt the method to classical versions of such systems.
On this aspect we must remark that some symmetry procedures usually
considered only for quantum systems can be extended in an appropriate
way to classical ones \cite{KN07}.


\section*{Acknowledgements.}

Partial financial support is acknowledged to Junta de Castilla y
Le\'on (Spain) under project VA013C05 and
the Ministerio de Educaci\'on y Ciencia of
Spain under project FIS2005-03989.


\end{document}